\documentclass[10pt,aps,prd,reprint,nofootinbib,floats,floatfix,amsfonts,amssymb,amsmath,preprintnumbers,notitlepage,superscriptaddress]{revtex4-1}

\usepackage{graphicx,color} 
\usepackage{hyphenat}
\usepackage{amsmath, amssymb, amsfonts,amsbsy,amsthm}
\usepackage{float}
\usepackage{subcaption}

\newcommand{\Kappa}[0]{\scalebox{1.5}{$\kappa$}}

\begin{document}

\begin{flushleft}
KCL-PH-TH/2021-71
\end{flushleft}

\title{Impact of Schumann resonances on the Einstein Telescope\\ and projections for the magnetic coupling function}

\author{Kamiel Janssens}
\affiliation{
Department of Physics, Universiteit Antwerpen, Antwerpen, Belgium}
\affiliation{Artemis, Université Côte d’Azur, Observatoire Côte d’Azur, Nice, France\\
}
\author{Katarina Martinovic}
\affiliation{Theoretical Particle Physics and Cosmology Group, \, Physics \, Department, \\ King's College London, \, University \, of London, \, Strand, \, London \, WC2R \, 2LS, \, UK}
\author{Nelson Christensen}
\affiliation{Artemis, Université Côte d’Azur, Observatoire Côte d’Azur, Nice, France\\
}
\author{Patrick M. Meyers}
\affiliation{School of Physics, University of Melbourne, Parkville, VIC 3010, Australia}
\affiliation{OzGrav, University of Melbourne, Parkville, VIC 3010, Australia}
\author{Mairi Sakellariadou}
\affiliation{Theoretical Particle Physics and Cosmology Group, \, Physics \, Department, \\ King's College London, \, University \, of London, \, Strand, \, London \, WC2R \, 2LS, \, UK}

\date{\today}

\begin{abstract}

Correlated magnetic noise in the form of Schumann resonances could introduce limitations to the gravitational-wave background searches of future Earth-based gravitational-wave detectors. We consider recorded magnetic activity at a candidate site for the Einstein Telescope, and forecast the necessary measures to ensure that magnetic contamination will not pose a threat to the science goals of this third-generation detector. In addition to global magnetic effects, we study local magnetic noise and the impact it might have on co-located interferometers. We express our results as upper limits on the coupling function of magnetic fields to the interferometer arms, implying that any larger values of magnetic coupling into the strain channel would lead to a reduction in the detectors' sensitivity. For gravitational-wave background searches below $\sim 30$ Hz it will be necessary for the Einstein Telescope magnetic isolation coupling to be two to four orders of magnitude better than that measured in the current Advanced LIGO and Virgo detectors.
\end{abstract}

\maketitle

\section{Introduction}
\label{sec:Introduction}

When searching for an isotropic gravitational-wave background (GWB)~\cite{Christensen_2018} one typically uses cross-correlation techniques between spatially separated interferometers to detect a correlated signal that is below the local noise of either individual instrument. However, globally correlated noise sources remain present in the cross-correlated data and can therefore affect the analysis. An example of such a source is the Schumann resonances~\cite{Schumann1,Schumann2}. Schumann resonances are extremely low frequency ($<$ 50Hz) electromagnetic excitations in the cavity formed by the Earth's surface and the ionosphere, driven by lightning strikes across the globe. Given their global character, the Schumann resonances are correlated over distances of several thousands of kilometers and longer.

Schumann resonances have magnetic strengths of the order of 1 pT at the fundamental mode, which has an observed frequency of 7.8 Hz. They are expected to couple magnetically to gravitational-wave (GW) interferometers via, e.g. the mirror suspension systems, electric cables and the electronics, thereby inducing a correlated signal of terrestrial origin \cite{Thrane:2013npa,Thrane:2014yza,galaxies8040082,Nguyen_2021}.

The impact of Schumann resonances on the ongoing searches for a GWB with the current generation of Earth-based interferometers (LIGO \cite{2015}, Virgo \cite{VIRGO:2014yos} and KAGRA \cite{PhysRevD.88.043007}) has been thoroughly investigated \cite{Thrane:2013npa,Thrane:2014yza,Coughlin:2016vor,Himemoto:2017gnw,Coughlin:2018tjc,Himemoto:2019iwd,Meyers:2020qrb}. The latest results in a search for an isotropic GWB showed that Schumann resonances are below the detector sensitivity; however as the sensitivity increases they could limit the search \cite{KAGRA:2021kbb}. More importantly, one must study their effect on searches with future Earth-based interferometers which aim to have approximately one order of magnitude improvement in sensitivity compared to current instruments.

In this study we focus on the European proposal for a third-generation (3G) GW interferometer: the Einstein Telescope (ET) \cite{Punturo:2010zz}. The current proposal consists of three interferometers with an opening angle $\pi$/3 forming an equilateral triangle. The no-longer operational Sos Enattos mine in Sardinia, Italy, is one of the possible locations to host the future ET interferometer, another being the Euregio Rhein-Maas at the intersection of the Belgian, Dutch and German borders \cite{Amann:2020jgo}. Please note that this paper does not contain a site comparison and typical magnetic spectra will be used to make statements on the impact of magnetic fields on ET, regardless of its exact location.

The research and development, and design phase of ET is ongoing. With this consideration we investigate the impact of fundamental magnetic noise sources, such as Schumann resonances, on the ET interferometers. We construct the maximal allowed magnetic coupling function so that the fundamental magnetic sources are not limiting the sensitivity of ET, either as an instrument itself or in its use for a search for an isotropic GWB.

In addition to low-frequency Schumann resonances, there are high-frequency correlated magnetic fields from individual lightning strikes~\cite{Kowalska-Leszczynska:2016low}, mainly situated in the frequency range 100 Hz - 1 kHz~\cite{ball:2020}. We will investigate the impact of these on ET's sensitivity, as well as consider possible limitations due to infrastructural noise.

The paper is organised as follows.
In Sec.~\ref{sec:Math} we discuss the key ingredients of the formalism of the search for an isotropic GWB and the way correlated magnetic noise can enter in the analysis. This formalism can be rewritten to allow us to calculate the maximal allowed magnetic coupling function to prevent magnetic contamination.
Sec.~\ref{sec:Data} focuses on the different magnetic data sets that will be used to make projections about the magnetic coupling function for ET. In Sec.~\ref{sec:Results} the magnetic coupling functions are constructed and we will discuss the impact of our results and their interpretation. A conclusion is presented in Sec.~\ref{sec:Conclusion}.

\section{Gravitational-Wave background}
\label{sec:Math}

The key figure of merit when searching for an isotropic GWB is
$\Omega_{\rm GW}(f)$~\cite{Christensen_2018,PhysRevD.46.5250,PhysRevD.59.102001,LivingRevRelativ20}:

\begin{equation}
    \label{eq:omgeaGWB}
    \Omega_{\rm GW}(f) = \frac{1}{\rho_{\rm c}}\frac{\text{d}\rho_{\rm GW}}{\text{d}\ln f}~,
\end{equation}
where we have defined the energy density, $\text{d}\rho_{\rm GW}$, contained in a logarithmic frequency interval, $\text{d}\ln f$, divided by the critical energy density $\rho_{\rm c} = 3H_0^2c^2/(8\pi G)$ for a flat Universe. $H_0$ is the Hubble constant, $c$ is the speed of light and $G$ is Newton's constant. We use the 15-year Planck value of 67.9 km s$^{-1}$ Mpc$^{-1}$ for $H_0$~\cite{Planck:2015fie}. 

To perform a search for a GWB, one typically uses a cross-correlation statistic that is an unbiased estimator for $\Omega_{\rm GW}(f)$\footnote{This is statement holds when one assumes the GWB is isotropic, Gaussian, stationary and unpolarized.}~\cite{PhysRevD.59.102001,LivingRevRelativ20}:
\begin{equation}
    \label{eq:cross-correlationstatistic}
\hat{C}_{IJ}(f) = \frac{2}{T_{\textrm{obs}}} \frac{{\rm{Re}}[\Tilde{s}^*_I(f)\Tilde{s}_J(f)]}{\gamma_{IJ}(f)S_0(f)}~,
\end{equation}
for interferometers $I$ and $J$, where
$\Tilde{s}_I(f)$ is the Fourier transform of the time domain strain data $s_I(t)$ measured by interferometer $I$, and $\gamma_{IJ}$ the normalized overlap reduction function (ORF) which encodes the baseline's geometry~\cite{PhysRevD.46.5250,Romano:2016dpx}. 
We approximate ET as three interferometers and ignore the details of the xylophone configuration~\cite{Hild:2010id}; this does not affect our results.

In what follows, we assume the three co-located, 10km long arm interferometers in triangle configuration of ET, namely ${\rm ET}_1, {\rm ET}_2, {\rm ET}_3$, as having identical sensitivity. Furthermore, we neglect the difference in $\gamma_{IJ}$ between the baseline pairs $IJ =  {\rm ET}_1{\rm ET}_2, {\rm ET}_1{\rm ET}_3, {\rm ET}_2{\rm ET}_3$, where we will be using $\gamma_{{\rm ET}_1{\rm ET}_2}$ from now on. For frequencies under 1 kHz the relative differences are~\cite{Andersson2009EinsteinTD}:  
$$\frac{|\gamma_{{\rm ET}_1{\rm ET}_2}-\gamma_{{\rm ET}_1{\rm ET}_3}|}{\gamma_{{\rm ET}_1{\rm ET}_2}}< 5 \times 10^{-7},$$  $$\frac{|\gamma_{{\rm ET}_1{\rm ET}_2}-\gamma_{{\rm ET}_2{\rm ET}_3}|}{\gamma_{{\rm ET}_1{\rm ET}_2}}< 2 \times 10^{-7},$$ justifying our choice to neglect the difference between baseline pairs.

The normalisation factor $S_0(f)$ is given by $S_0(f)=(9H_0^2)/(40\pi^2f^3)$ and $T_{\textrm{obs}}$ is the total observation time of the data-collecting period \footnote{The form of $S_0(f)$ for ET differs from that one of e.g. LIGO by a factor of 3/4. This is due to the difference in opening angle between interferometers' arms ($\pi$/2 for LIGO and $\pi$/3 for ET) that leads to different normalization factors in the ORF of LIGO and ET baseline pairs \cite{Romano:2016dpx}.}. In the absence of correlated noise this cross-correlation statistic is an unbiased estimator of $\Omega_{\rm GW}(f)$.

Equivalent to the cross-correlation statistic, one can construct a magnetic cross-correlation statistic~\cite{Thrane:2013npa,Thrane:2014yza} 
\begin{equation}
    \label{eq:C_Mag}
    \begin{aligned}
        \hat{C}_{{\rm mag},{\rm ET}_1{\rm ET}_2}(f) &= |\scalebox{1.5}{$\kappa$}_{\rm ET}(f)|^2 M_{{\rm ET}_1{\rm ET}_2}, \\
\mbox{where~~}      M_{{\rm ET}_1{\rm ET}_2} &= \frac{2}{T_{\textrm{obs}}} \frac{|\Tilde{m}^*_{{\rm ET}_1}(f)\Tilde{m}_{{\rm ET}_2}(f)|}{\gamma_{{\rm ET}_1{\rm ET}_2}(f)S_0(f)}~,
    \end{aligned}
\end{equation}
and
$\scalebox{1.5}{$\kappa$}_{\rm ET}(f)$ describes the coupling from magnetic fields to interferometer ${\rm ET}_1$, where we have used $\scalebox{1.5}{$\kappa$}_{\rm ET}(f)=\scalebox{1.5}{$\kappa$}_{{\rm ET}_1}(f)=\scalebox{1.5}{$\kappa$}_{{\rm ET}_2}(f)$. We denote by $\Tilde{m}_{{\rm ET}_1}(f)$ the Fourier transform of the time domain data $m_{{\rm ET}_1}(t)$ measured by a magnetometer at site ${\rm ET}_1$, and by $T_{\textrm{obs}}$ the duration of the segments used when Fourier transforming the magnetic data\footnote{Note the different meaning of $T_{\textrm{obs}}$ in Eq.~(\ref{eq:cross-correlationstatistic}) and Eq.~(\ref{eq:C_Mag}), since in Eq.~(\ref{eq:cross-correlationstatistic}) the full data set is analysed in one segment, whereas in Eq.~(\ref{eq:C_Mag}) we divide the data in multiple segments after which they are combined.}.
To construct a conservative magnetic cross-correlation statistic we take the modulus of $\Tilde{m}^*_{{\rm ET}_1}(f)\Tilde{m}_{{\rm ET}_2}(f)$ rather than taking only the real part into account~\cite{LIGOScientific:2019vic,KAGRA:2021kbb}.

When analysing data in a search for an isotropic GWB using interferometers $I$ and $J$, one typically constructs the magnetic cross-correlation statistic $\hat{C}_{{\rm mag},IJ}(f)$ to investigate if the observed magnetic fields might result in correlated noise in the analysis \cite{LIGOScientific:2019vic,KAGRA:2021kbb}. The magnetic coupling functions $\scalebox{1.5}{$\kappa$}_{I,J}(f)$ are measured by injecting magnetic fields with known amplitude and frequency and observing their impact on the output in the GW data $s_{I,J}(t)$~\cite{galaxies8040082,Nguyen_2021}.

However, in this study we will consider a different approach: given a desired sensitivity and a magnetic spectrum $|\Tilde{m}^*_{{\rm ET}_1}(f)\Tilde{m}_{{\rm ET}_2}(f)|$, we estimate the maximal allowed magnetic coupling function $\scalebox{1.5}{$\kappa$}_{\rm ET}(f)$ such that ET will not be limited by this magnetic noise.

The sensitivity for a GWB search is different to the instantaneous sensitivity of the ET interferometer, referred to as the one-sided amplitude spectral density (ASD) $P_{\rm ET}(f)=P_{{\rm ET}_1}(f)=P_{{\rm ET}_2}(f)=P_{{\rm ET}_3}(f)$. In the case of an isotropic GWB search, the sensitivity is given by the standard deviation on the cross-correlation statistic defined in Eq.~\ref{eq:cross-correlationstatistic}. In the small signal-to-noise ratio (SNR) limit, this is given by~\cite{PhysRevD.46.5250,PhysRevD.59.102001,LivingRevRelativ20} 
\begin{equation}
    \label{eq:sigmaGWB}
    \sigma_{{\rm ET}_1{\rm ET}_2}(f) \approx \sqrt{\frac{1}{2T_{\rm obs}\Delta f}\frac{P_{\rm ET}^2(f)}{\gamma_{{\rm ET}_1{\rm ET}_2}^2(f)S_0^2(f)}}~,
\end{equation}
where $\Delta f$ is the frequency resolution, and $\sigma_{{\rm ET}_1{\rm ET}_2}(f)$ defines our uncertainty and therefore our sensitivity in a single frequency bin used in the analysis. However, when searching for an isotropic GWB, one typically expects a broadband signal and often assumes a power-law GWB model. Hence, a useful measure of GWB sensitivity is also the power-law integrated (PI) curve. The PI curve, $\Omega^{\rm PI}_{{\rm ET}_1{\rm ET}_2}(f)$ is constructed using $\sigma_{{\rm ET}_1{\rm ET}_2}(f)$ so that its tangent at any frequency represents the sensitivity at which one could detect a power-law $\Omega_{\rm GW}(f)$ with an SNR of 1 for the ${\rm ET}_1{\rm ET}_2$ baseline~\cite{Thrane:2013oya}. Therefore, it serves as a good figure of merit to identify broadband noise sources that could limit the sensitivity of a GWB search.

To ensure that magnetic noise does not obstruct the isotropic search for a GWB, we construct an upper limit on the magnetic coupling function that we label  ``GWB," in the following way. We use Eq.~(\ref{eq:C_Mag}) and take the upper limit for the magnetic cross-correlation $ \hat{C}_{{\rm mag},{\rm ET}_1{\rm ET}_2}(f)$ to be the 1$\sigma$-PI sensitivity curve -- $\Omega^{\rm PI}_{{\rm ET}_1{\rm ET}_2}$ -- after one year of taking data:
\begin{equation}
    \label{eq:CFStochastic}
    \scalebox{1.5}{$\kappa$}_{\rm ET}^{ \rm GWB}(f) \equiv \sqrt{ \frac{\Omega^{\rm PI}_{{\rm ET}_1{\rm ET}_2}}{M_{{\rm ET}_1{\rm ET}_2}}}~.
\end{equation}

In addition, we explore a complimentary method for computing upper limits of the magnetic coupling function $\scalebox{1.5}{$\kappa$}_{\rm ET}(f)$. To investigate the impact of magnetic noise sources on the ASD of an individual interferometer, we construct what we refer to as the
``ASD'' upper limit: 
\begin{equation}
    \label{eq:CFInfrastructure}
    \scalebox{1.5}{$\kappa$}_{\rm ET}^{ \rm ASD}(f) \equiv k~ 
    \frac{P_{\rm ET}(f)}{P_{\rm mag}(f)}~,
    \end{equation}
where $k$ is set to be $1/10$ to require any single technical noise contribution to be a factor of 10 lower than ET's ASD, $P_{\rm ET}(f)$. $P_{\rm mag}(f)$ is the one-sided ASD of the magnetometers witnessing the local noise of ET, assuming the magnetic noise to be the same for the three individual ET interferometers ${\rm ET}_1$, $\rm ET_2$ and $\rm ET_3$.
If magnetic fields couple significantly at this level, they will limit  the expected sensitivity of the interferometer. The  ``ASD'' upper limit should be investigated to prevent magnetic noise from drastically impacting \textit{all} science goals of the GW interferometer.

To make full use of ET's capabilities when searching for an isotropic GWB, one should use  
$\scalebox{1.5}{$\kappa$}_{\rm ET}^{ \rm GWB}(f)$.
On the contrary, to find the level at which magnetic fields might directly impact the instantaneous sensitivity achieved by ET, then $\scalebox{1.5}{$\kappa$}_{\rm ET}^{ \rm ASD}(f)$ is the relevant measure.\\

\section{Data}
\label{sec:Data}

In this analysis we use observed, rather than simulated, magnetic data. However, since the location and exact positioning of ET is unknown we will estimate $|\Tilde{m}^*_{{\rm ET}_1}(f)\Tilde{m}_{{\rm ET}_2}(f)|$ using a variety of observed magnetic spectra that we describe below.

Schumann resonances have a relatively similar amplitude regardless where on Earth they are measured, with variations of a factor of 2-3 in some cases~\cite{Coughlin:2018tjc}. The amplitude of  the fundamental mode ($\sim$ 8 Hz) rarely exceeds 1 pT\footnote{See Fig. 1 of~\cite{Coughlin:2018tjc} for a comparison of Schumann resonance measurements across the globe.}. Moreover, daily and seasonal variations of the resonances' amplitude occur.

As a part of site characterisation of the Sos Enattos mine for ET~\cite{Naticchioni:2020kfb,RRomero}, magnetic field measurements were taken from 5Hz to 100Hz in the exceptionally quiet environment inside the unused mine. We use 48 days of data taken from Nov 14 2019 to Dec 31 2019, using a single-axis Metronix MFS-06e magnetometer positioned inside the Sos Enattos mine about 200 m below ground level. The magnetometer is sampled at 250 Hz.
Given these on-site measurements are available for Sos Enattos but not for the Euregio Rhein-Maas ET-candidate site, we will use the magnetic spectrum observed at Sos Enattos as a reference estimate for the magnetic spectrum at the future ET site: $\Tilde{m}_{{\rm ET}}(f) = 2 \Tilde{m}_{\rm Sos Enattos}(f)$. The factor 2 allows for higher magnetic fields at the final site and to be more conservative.

Further investigations into magnetic noise at Euregio Rhein-Maas could help compare the two currently proposed sites, but we do not make any statement about which site will be used as the actual location of the ET detector. Site-specific amplifying or reducing ambient magnetic fields are typically a second order effect that can be ignored. However, as shown by measurements at KAGRA, local amplification of ambient magnetic fields may not always be negligible~\cite{Atsuta_2016,KAGRA:2017tir}.

\begin{figure}
    \centering
    \includegraphics[width=\linewidth]{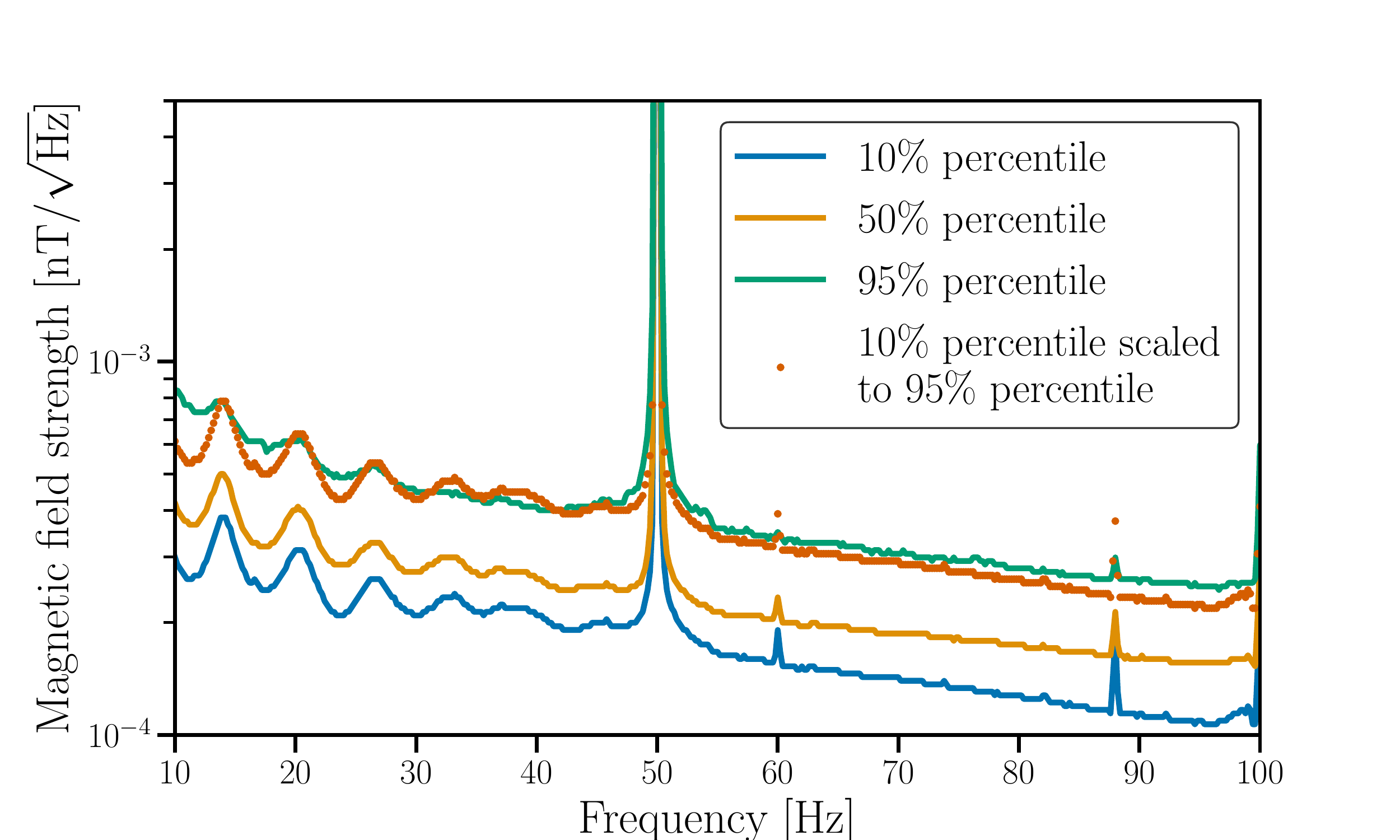}
    \caption{Sos Enattos magnetic ASD constructed using 48 days of data from Nov 14 2019 to Dec 31 2019. A one-directional magnetometer was employed to collect the data in the mine approximately 200 m below ground level. The line at 50Hz is coming from the power mains.}
    \label{fig:SosEnnatosPSD}
\end{figure}

We show in Fig.~\ref{fig:SosEnnatosPSD}  percentiles of the magnetic ASD as measured at Sos Enattos. We use the shape of the 10\% magnetic percentile curve, since it captures the peaks of the Schumann spectrum very well. 
We then scale the amplitude to match the amplitude of the 95\%-percentile curve at 7.8 Hz, the first Schumann mode. This gives a good prediction for the Schumann resonance spectrum at Sos Enattos. 

This low-frequency range is where most of the sensitivity is for the ongoing isotropic GWB searches, with the most recent LIGO-Virgo observing run containing 99\% sensitivity to a flat $\Omega_{\rm GW}$ spectrum below 100 Hz \cite{KAGRA:2021kbb}. The sensitivity at high frequencies is suppressed since the ORF for the LIGO and Virgo baseline pairs drops significantly with frequency and approaches zero at a few hundreds of Hz. The same, however, may not be true for the ET detectors, whose geometry leads to no suppression at high frequency, and the anticipated ORF remains approximately constant up to 1 kHz, e.g. $\gamma_{{\rm ET}_1{\rm ET}_2}$(1 kHz)/$\gamma_{{\rm ET}_1{\rm ET}_2}$(1 Hz) = 99.48\%. We therefore look to place upper limits on the magnetic coupling function at high frequencies.

We estimate the correlated magnetic noise spectrum above 100Hz by computing the cross spectral density (CSD) between magnetometers at LIGO Hanford and LIGO Livingston. The separation of several thousands of kilometers between the two sites ensures that the estimated spectrum only contains fundamental, global effects rather than local contributions. We consider the Hanford-Livingston baseline as the most conservative choice for magnetic CSD since these are the closest interferometers and the impact of individual lightning strikes at frequencies above $\sim$ 100 Hz drastically attenuates with increasing distance. Hanford and Livingston both have two low-noise magnetometers on site, positioned at low-noise locations orientated along the interferometers arms. We make use of an ``omni-directional'' magnetic CSD, where we take into account all possible cross-correlation combinations between the magnetometer pairs at both interferometer sites, 
\begin{equation}
    \label{eq:OmnidirectionalCSD}
    \begin{aligned}
    {\rm CSD}_{\rm HL} =& \left[ \right.  |\Tilde{m}^*_{H_x}(f)\Tilde{m}_{L_x}(f)| ^2 + |\Tilde{m}^*_{H_x}(f)\Tilde{m}_{L_y}(f)|^2  \\
                     &+ |\Tilde{m}^*_{H_y}(f)\Tilde{m}_{L_x}(f)| ^2 +  |\Tilde{m}^*_{H_y}(f)\Tilde{m}_{L_y}(f)|^2 \left.\right] ^{1/2}~.\\
    \end{aligned}
\end{equation}
Here $H_x,H_y$ ($L_x,L_y$) represent the two orthogonal magnetometers at Hanford (Livingston) pointing along the interferometer's $x$- and $y$-arms. We use data recorded from Apr 2 2019 to Mar 27 2020, which approximately matches the LIGO-Virgo third observing run, named O3.

We assume ${\rm CSD}_{\rm HL}$ is a realistic estimation for the magnetic noise at ET, where we include an additional factor of 2 reflecting our uncertainty on the magnetic spectrum \footnote{A factor of 2 at the level of individual magnetic fields $\Tilde{m}_{{\rm ET}}(f)$, corresponds to a factor of $2^2$ at the level of $|\Tilde{m}^*_{{\rm ET}_1}(f)\Tilde{m}_{{\rm ET}_2}(f)|$.}:     $\frac{2}{T_{obs}}|\Tilde{m}^*_{{\rm ET}_1}(f)\Tilde{m}_{{\rm ET}_2}(f)|= 4 {\rm CSD}_{\rm HL}$.

We also investigate the impact of local magnetic noise. One should study local magnetic noise carefully as this could be a correlated noise source as well because each local corner station for the triangular ET set-up will likely house mirrors for two of the three interferometers (separated by $\sim 300$ m and in different vacuum tubes). 
To model local magnetic fields of a GW interferometer, we consider the local magnetic noise measured in the central building (CEB) at Virgo site~\cite{galaxies8040082}. We use the 90\% magnetic percentile of data collected between Feb 10 2020 and Feb 16 2020 - during the second half of the third observing run, O3b. This spectrum is quite similar to the one observed in the Virgo CEB during O2~\cite{Cirone_2019}. 

We assume the local noise at Virgo is a realistic estimation for the local magnetic noise at ET, where we include an additional factor of 2 reflecting our uncertainty on the magnetic spectrum:     $\Tilde{m}_{{\rm ET}_1}(f)= 2 \Tilde{m}_{\rm V_{CEB}}(f)$.

\section{Results}
\label{sec:Results}

\begin{figure}
    \centering
    \includegraphics[width=\linewidth]{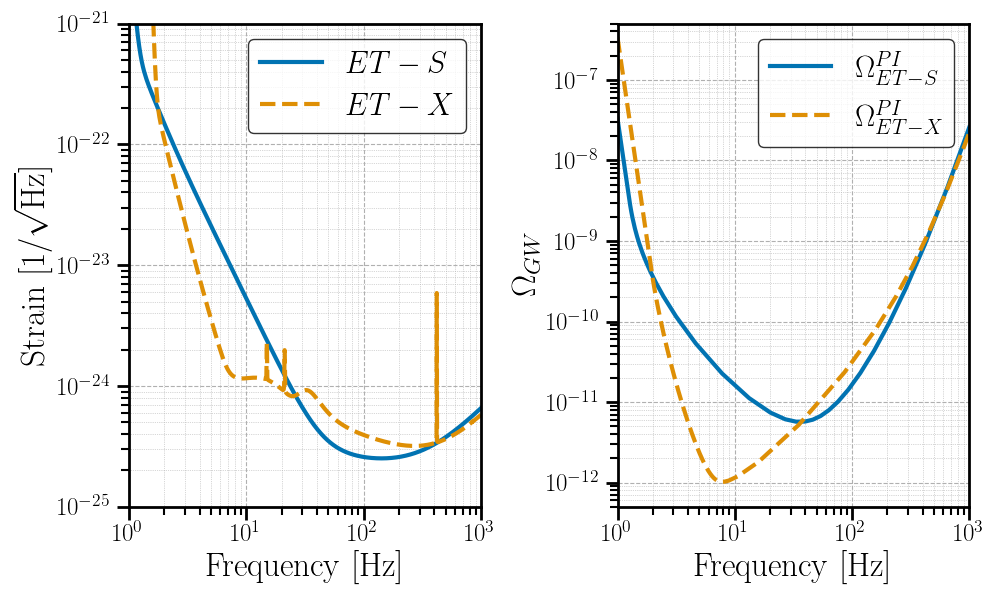}
    \caption{The ET configurations -- ET-S, ET-X~\cite{Hild_2009,Hild:2010id} -- and their anticipated sensitivity curves (left panel) as well as their power-law integrated curves after a year-long observation (right panel).}
    \label{fig:et-sensitivity}
\end{figure}

We construct two complementary measures---$\scalebox{1.5}{$\kappa$}_{\rm ET}^{ \rm GWB}(f)$ and $\scalebox{1.5}{$\kappa$}_{\rm ET}^{ \rm ASD}(f)$---of the magnetic coupling function. 
In our analysis, we use two preliminary design studies, called $\rm ET_{\rm Single}$ and $\rm ET_{\rm Xylophone}$\footnote{In previous literature $\rm ET_{\rm Single}$ and $\rm ET_{\rm Xylophone}$ have also been referred to as respectively ET-B, ET-D.}~\cite{Hild_2009,Hild:2010id}. In the remainder of this paper we will use the notation ET-S and ET-X as abbreviations to refer to $\rm ET_{\rm Single}$, respectively $\rm ET_{\rm Xylophone}$. In Fig.~\ref{fig:et-sensitivity} we show the sensitivity curves for both of these design options in the left panel. In the right panel we show the PI curves for each of these design options for 1 year of integration time. The ET-X configuration is an order of magnitude more sensitive at low frequencies as compared to ET-S. Fig.~\ref{fig:et-sensitivity}, shows ET-X could be sensitive to a GWB at the $\Omega_{\rm GW}(f) \sim 10^{-12}$ range, with an SNR=1 after 1 year of observation. This seems to be consistent with earlier investigations stating a GWB with strength $\Omega_{\rm GW}(f) = 2\cdot 10^{-12}$ would be detected with an SNR=5 after 1.3 years of observation time\cite{PhysRevD.102.063009}.

The upper limits presented here assume the reduction of the magnetic coupling is the only pursued method to prevent magnetic fields from coupling significantly to the interferometer. Methods such, as the use of Wiener filters~\cite{Coughlin:2018tjc}, could also be used to reduce the effects of correlated magnetic noise. However, the best strategy for ET will be to design the magnetic isolation to be as good as possible.

\begin{figure}
    \centering
    \includegraphics[width=\linewidth]{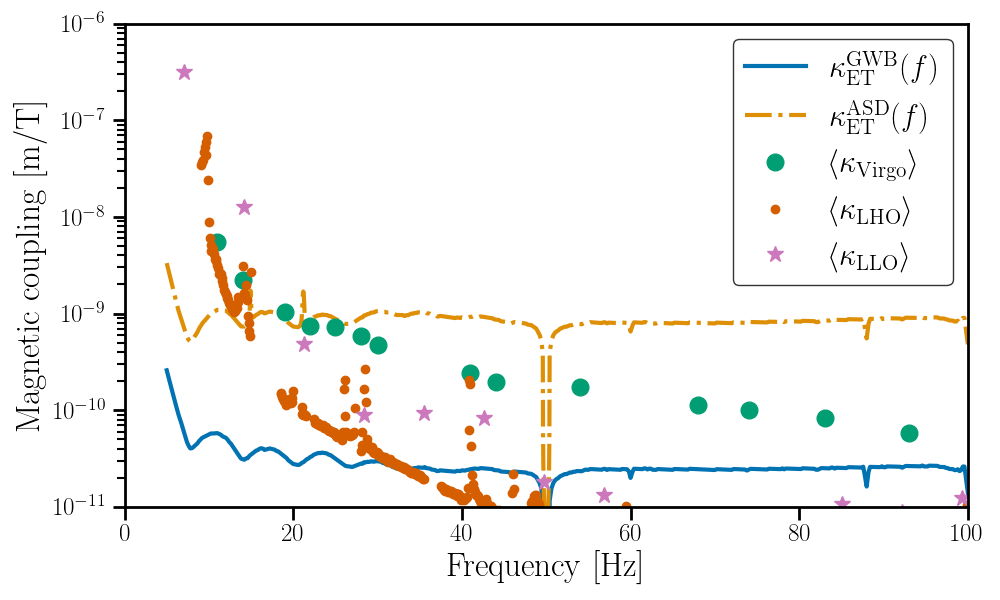}
    \caption{``ASD'' and ``GWB'' magnetic coupling function upper limits for $\rm ET-X$ design sensitivity. Also included are the average of the measurements of the coupling functions at LIGO Hanford, LIGO Livingston and Virgo during the O3 run for comparison.}
    \label{fig:et-c}
\end{figure}

\begin{figure}
    \centering
    \includegraphics[width=\linewidth]{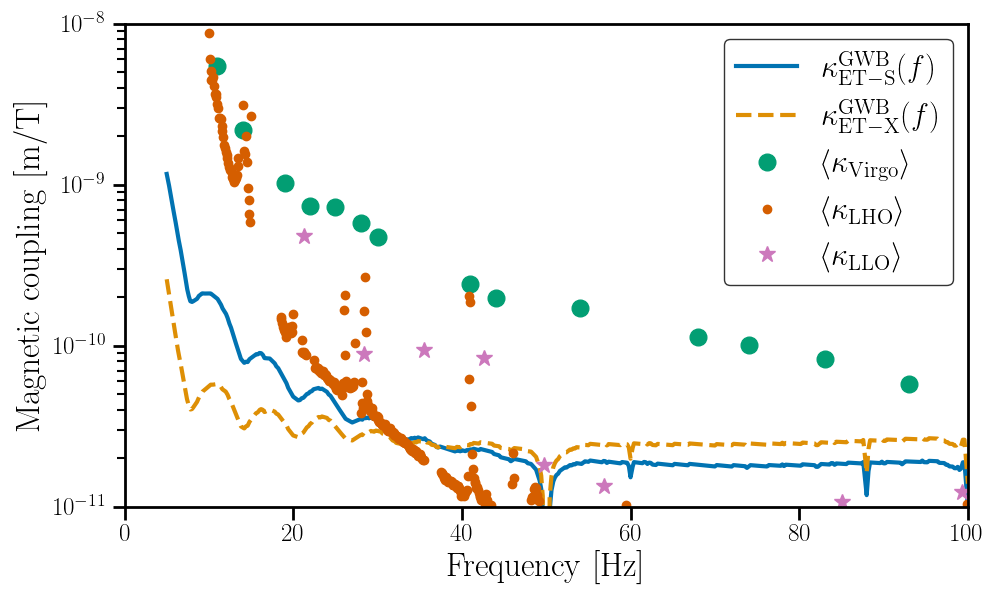}
    \caption{Variation in the ``GWB'' magnetic coupling function upper limits for the different ET designs. Also included are the average of the measurements of coupling functions at LIGO Hanford, LIGO Livingston and Virgo during the O3 run for comparison.}
    \label{fig:et-bcd}
\end{figure}

\begin{figure}
    \centering
    \includegraphics[width=\linewidth]{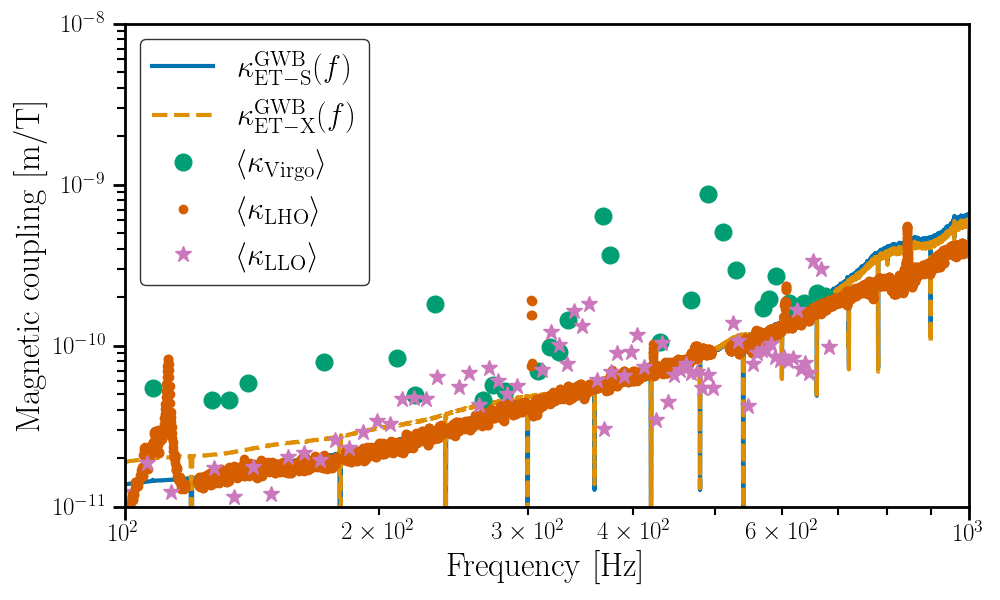}
    \caption{Upper limits on ``GWB'' magnetic coupling function of ET-S and -X at high frequencies. Also included is the average of the measurements of coupling functions at LIGO Hanford, LIGO Livingston and Virgo during the O3 run for comparison.}
    \label{fig:high-f}
\end{figure}

In Fig.~\ref{fig:et-c} we show limits on the coupling imposed by our target sensitivity measures. The $\Kappa^{\textrm{GWB}}_{\rm ET-X}(f)$ limits are indicated by the solid blue curve, while $\Kappa^{\textrm{ASD}}_{\rm ET-X}(f)$ limits are denoted by the dash-dotted yellow curve. The average coupling measurements made at Virgo~\cite{galaxies8040082,Cirone_2018}, LIGO Hanford and Livingston~\cite{Davis_2021, alog:mag_fac,alog:O3PEMInjections, alog:WeeklyMagneticInjections} during the O3 run are indicated by the green circles, smaller orange circles, and magenta stars respectively. Please note that the magnetic coupling measurements only start from 11Hz, 9Hz and 7Hz for respectively the Virgo, LIGO Hanford and LIGO Livingston interferometers. In Fig.~\ref{fig:et-bcd} and Fig.~\ref{fig:high-f} we show high and low-frequency coupling limits, $\Kappa^{\textrm{GWB}}_{\textrm{ET}}(f)$, respectively for each of the design options (solid blue for ET-S, yellow dashed for $\rm ET-X$). The average coupling measurements at Virgo, Hanford, and Livingston are the same as in Fig.~\ref{fig:et-c}. 

We notice that the upper limit on the magnetic coupling at certain frequencies is allowed to be greater than the magnetic coupling at current observatories\footnote{Above 100 Hz the weekly measurements of the magnetic coupling are often upper limits rather than an actual measurements. However during some periods of ``extreme'' magnetic coupling such high values are actually measured as well~\cite{galaxies8040082}.}, for example $\gtrsim 30$~Hz in the $\Kappa^{\textrm{ASD}}_{\rm ET-X}(f)$ curve in Fig.~\ref{fig:et-c}. This means that one can be less concerned about Schumann resonance magnetic noise coupling into ET as compared to LIGO/Virgo detectors. 
The reason for this result can be seen by considering the difference between the magnetic coupling measured in units of [$\rm {T^{-1}}$] \footnote{The unit of magnetic coupling function [$\rm {T^{-1}}$] is often referred to as [strain $\rm {T^{-1}}$].} and in units of [$\rm {m~T^{-1}}$]. The latter takes into account the arm-length of the interferometer, $\textit{L}_{\rm arm}$, 
\begin{equation*}
    \label{eq:TF_strain}
    \scalebox{1.5}{$\kappa$}[\rm {T^{-1}}] =  \frac{\scalebox{1.5}{$\kappa$}\,[\rm {m~ T^{-1}}]}{\textit{L}_{\rm arm}[m]}~.
\end{equation*}
Since ET is planned to have a 10 km arm-length, instead of 4 km (LIGO) or 3 km (Virgo), the test masses displacements due to magnetic effects measured in the units [$\rm {m~T^{-1}}$] is allowed to be larger compared to existing interferometers.\\ 

We emphasise that Fig.~\ref{fig:et-c} and Fig.~\ref{fig:et-bcd} use data measured at the Sos Enattos candidate site for ET. However, given the similar amplitude of the Schumann resonances around the globe, these results could be transferable to another location and more specifically a second candidate site in the Euregio Rhein-Maas. That is assuming there is no extreme magnification of magnetic fields due to local effects, as observed at KAGRA~\cite{Atsuta_2016,KAGRA:2017tir}. To account for such a possibility we introduced a factor of 1/2 in our estimates of the coupling upper limits.

We illustrate in Fig.~\ref{fig:high-f} the effect of magnetic fields at ET above a 100 Hz. An example of an analysis that could target a signal at these higher frequencies is the study of a GWB from unresolved millisecond pulsars. This search is complementary to the standard continuous-wave search of individual pulsars, and can help constrain ellipticity of rotating neutron stars~\cite{Talukder:2014eba,LIGOScientific:2020gml, LIGOScientific:2021ozr}. Current forecasts predict an improvement of one to two orders of magnitude in the sensitivity to ellipticity going from LIGO-Virgo to ET. However, correlated magnetic noise at high frequencies could weaken the ellipticity constraints (see, Fig.~6 in~\cite{Maggiore:2019uih}), and should thus be treated carefully.

In our high-frequency analysis we have used the magnetic CSD measured between LIGO Hanford and LIGO Livingston, two widely separated sites, to ensure we are not dominated by local effects. However, since magnetic fields from the sources contributing to this magnetic noise (e.g. individual lightning strikes~\cite{ball:2020}) are attenuated over long distances, the fundamental magnetic spectrum at ET could be stronger compared to our predictions. 
Therefore one should be cautious interpreting the upper limits presented in Fig.~\ref{fig:high-f}.
For the magnetic coupling above 100Hz, we do not show the upper limits calculated using the ``ASD" formalism, Eq.~(\ref{eq:CFInfrastructure}), since these limits are less stringent than the measured magnetic coupling at present day interferometers. \\

\begin{figure*}[t]
    \centering
    \begin{subfigure}[t]{0.49\textwidth}
        \centering
        \includegraphics[width=\linewidth]{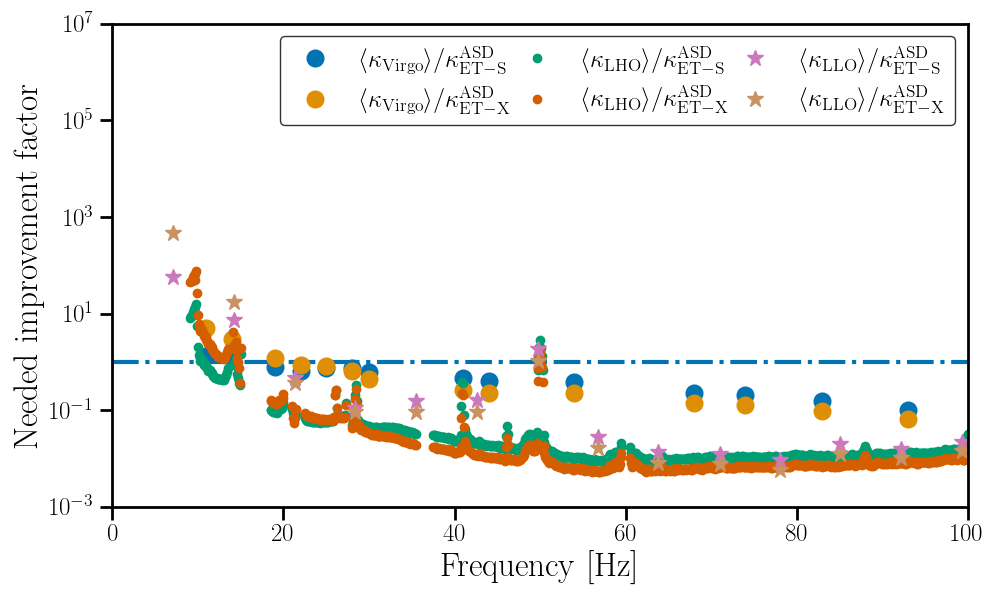} 
    \end{subfigure}
    \hfill
    \begin{subfigure}[t]{0.49\textwidth}
        \centering
        \includegraphics[width=\linewidth]{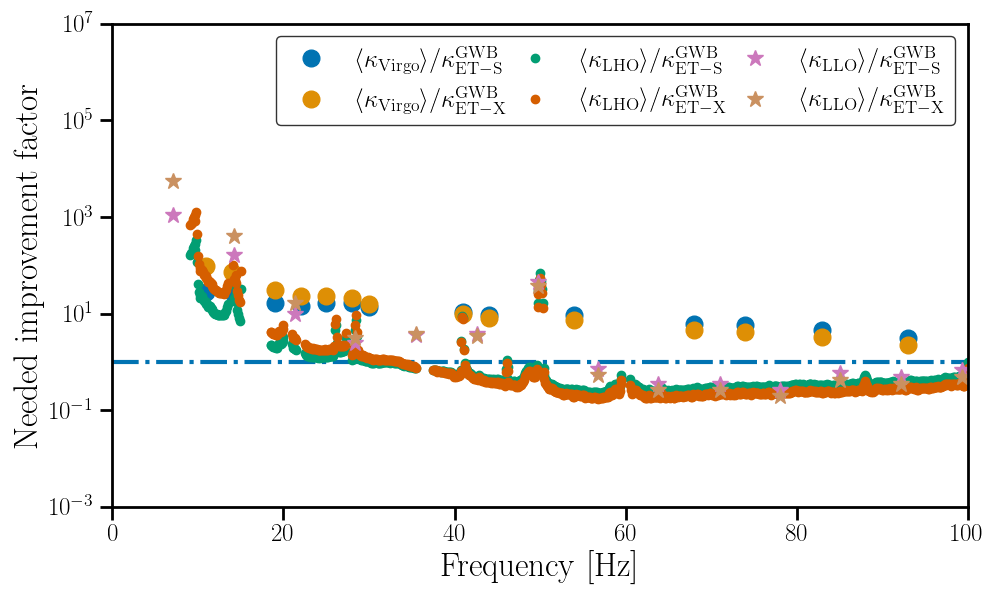} 
    \end{subfigure}
    \begin{subfigure}[t]{0.49\textwidth}
        \centering
        \includegraphics[width=\linewidth]{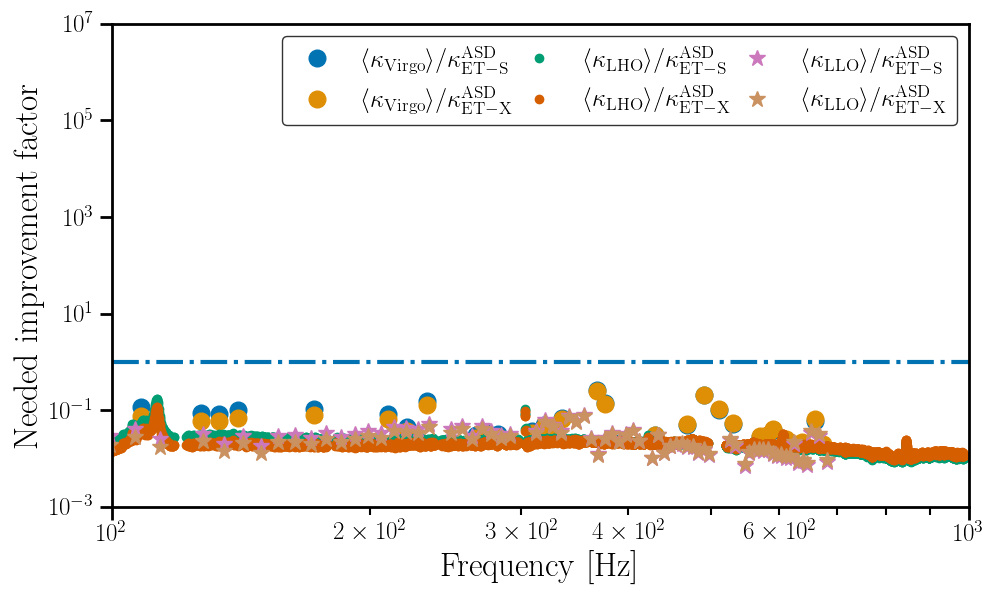} 
    \end{subfigure}
    \hfill
    \begin{subfigure}[t]{0.49\textwidth}
        \centering 
        \includegraphics[width=\linewidth]{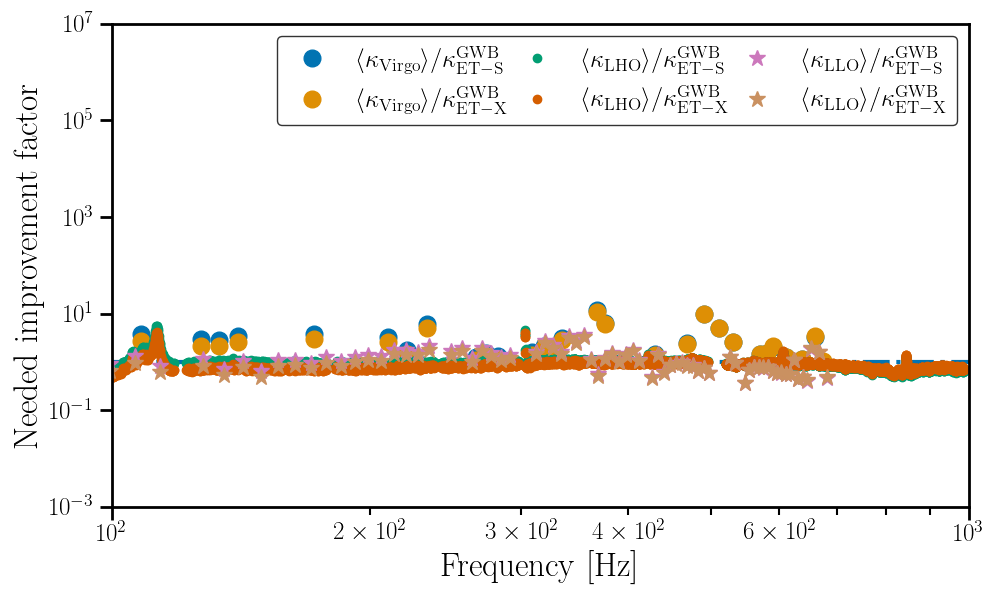}
    \end{subfigure}
    \caption{Needed improvement factor as a function of frequency for the ``ASD'' (left panels) and ``GWB'' (right panels) upper limits on the ET magnetic coupling function. The low-frequency (top panels) magnetic coupling poses a greater challenge for the operation of ET compared to the high-frequency (bottom panels) magnetic coupling, while ``GWB'' upper limits on the magnetic coupling are more constraining than the ``ASD'' ones. In all panels, the dash-dotted blue line indicates the line where no improvement is necessary.}
    \label{fig:ratios}
\end{figure*}

We summarise these results in Fig.~\ref{fig:ratios}, where we express the estimated ``ASD'' and ``GWB'' upper limits as a factor of improvement needed in the ET coupling function relative to Hanford, Livingston and Virgo coupling functions. We show how this factor varies with frequency, as well as how it changes with the choice of ET design sensitivity. For frequencies below $\sim 30$ Hz, the magnetic coupling would need to drastically reduced, with factors of improvement of the order $10^2-10^4$ needed for the isotropic GWB search, see top right panel. High-frequency coupling, on the other hand, would only require up to a factor of 10 reduction in magnetic coupling to run a successful GWB search (bottom right panel). Meanwhile, from the ``ASD'' upper limits we do not require any improvement in coupling function at frequencies above 30 Hz, see two left panels, confirming that this is a less conservative constraint on the magnetic coupling.

Note that above 30 Hz Advanced LIGO's magnetic coupling is dominated by induction of currents in cables~\cite{Nguyen_2021}. One mitigation strategy that could be followed in this scenario is using, as much as possible, a cabling network of optical fibers. The implementation of a large-scale optical fiber network has been investigated and implemented at CERN~\cite{SHOAIE201869}. An important factor for the reduced magnetic coupling for Advanced LIGO compared to Advanced Virgo is that LIGO uses an electrostatic test mass actuators whereas Virgo uses magnetic actuation~\cite{Nguyen_2021,Cirone_2018}. Further reducing the number of magnets attached to the suspensions should reduce magnetic coupling. Additional magnetic shielding can be a complementary method to reduce the magnetic coupling~\cite{Cirone_2019}.
Ultimately, if methods for magnetic coupling reduction are insufficient one could consider the cancellation of magnetic noise, similar to what is considered in the context of Newtonian Noise~\cite{PhysRevD.92.022001,Badaracco_2019}, albeit using magnetometers instead of seismometers.

\begin{figure*}
    \centering
    \includegraphics[width=0.8\linewidth]{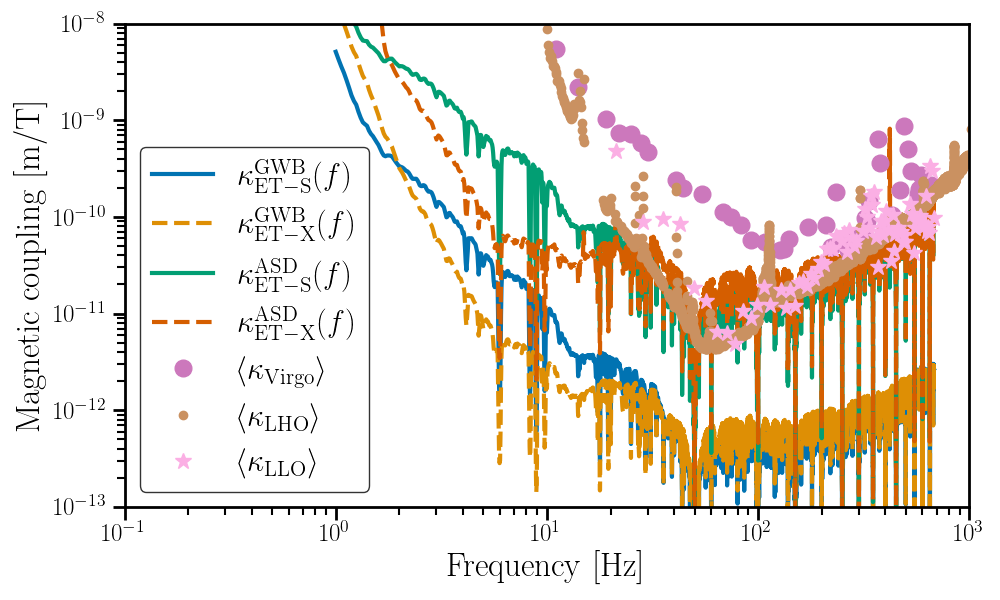}
    \caption{``ASD'' and ``GWB'' magnetic coupling function upper limits of all ET design sensitivities in the case the local magnetic noise is the same level as the CEB at Virgo during O3. Also included are the average of the measurements of coupling functions at LIGO Hanford, LIGO Livingston and Virgo during the O3 run for comparison.}
    \label{fig:local}
\end{figure*}

Finally, if one is unable to reduce the effect from local magnetic fields originating from e.g. used infrastructure, local magnetic noise sources will dominate the fundamental magnetic noise discussed above. This leads to the most stringent coupling upper limits, reported in Fig.~\ref{fig:local}. To construct these upper limits the magnetic noise as observed in the Virgo central building is used. This represents a realistic magnetic environment in present-day interferometers, however it may not be the most conservative. 

In the case of Virgo this local noise does not pose a serious problem in the GWB search since it is uncorrelated with local magnetic noise at far-away Hanford and Livingston detectors. Between co-located ET interferometers, however, the local noise could become correlated.     
This can lead to drastically more stringent upper limits on the magnetic coupling. If one wants to fully utilize the data for GWB searches, the magnetic coupling should be well below the magnetic coupling measured at Hanford, Livingston and Virgo, see Fig.~\ref{fig:local}.
Below $\sim$ 30 Hz, the instantaneous detector sensitivity will be limited by magnetic noise if the coupling is not reduced below the current day magnetic coupling of Hanford, which is already significantly smaller compared to the coupling measured at Livingston and Virgo.\\

\section{Conclusion}
\label{sec:Conclusion}

ET is a powerful and promising instrument for detecting a GWB, with an unprecedented low-frequency sensitivity compared to LIGO and Virgo~\cite{Maggiore:2019uih}. Methods have been proposed to separate an astrophysical GWB (from compact binary mergers) from a cosmological one~\cite{Regimbau:2016ike,Sachdev:2020bkk,Martinovic:2020hru}. 
ET could be sensitive at the SNR=1 level to a cosmological GWB $\Omega_{\rm GW}(f) \sim 10^{-12}$, but this depends critically on the low-frequency performance of ET. The ability to detect a GWB through correlation methods between multiple detectors assumes the absence of correlated noise~\cite{Christensen_2018}, however globally coherent magnetic fields have been identified as a limiting noise source for the present GW detector network~\cite{Thrane:2013npa,Thrane:2014yza,Meyers:2020qrb}. As we have shown, this is also the case for ET. 
More precisely, we have shown that the magnetic coupling functions for ET must be better than those of LIGO and Virgo by a factor of $10^2-10^4$ for frequencies below 30 Hz, in order to avoid correlated noise from Schumann resonances affecting GWB searches.

Reducing the magnetic coupling to prevent a significant impact on the interferometers, and also to ensure that local magnetic noise is as small as possible, is the best strategy for ET. This could be achieved by reducing the number of magnets attached to the suspensions \cite{Nguyen_2021}, additional shielding \cite{Cirone_2019}, and using optical fibers as much as possible for signal transmission \cite{Nguyen_2021}. There could also be a synergy with noise subtraction methods, such as Wiener filters~\cite{Coughlin:2018tjc}. Such methods have been investigated to cancel Newtonian Noise~\cite{PhysRevD.92.022001,Badaracco_2019}. However they also could be used to reduce the effects of correlated magnetic noise and loosen the requirements on the magnetic coupling, as presented here. 

Not reaching the reported upper limits on the magnetic coupling functions could have a direct impact on the search for a GWB with ET. Note, however, that reaching these upper limits is not necessarily a guarantee that there are no effects by magnetic fields on the search for a GWB. The importance of correlated magnetic noise coupling will need to be considered as ET is designed and constructed.

\acknowledgements
The authors acknowledge access to computational resources provided by the LIGO Laboratory supported by National Science Foundation Grants PHY-0757058 and PHY-0823459. 
Furthermore the authors acknowledge the Sos Enattos former mine, hosting the instrumentation used for this study, the Istituto Nazionale di Fisica Nucleare (INFN), the Istituto
Nazionale di Geofisica e Vulcanologia (INGV), the European Gravitational Observatory (EGO), the University of Sassary and the Regione Autonoma Sardegna for the support to the site characterization activities.

Furthermore the authors would like to thank Michael Coughlin, Irene Fiori and Jan Harms for useful comments and discussions. 

This paper has been given LIGO DCC number P2100356, Virgo TDS number VIR-1050A-21 and ET TDS number ET-0427A-21.

K.J. is supported by FWO-Vlaanderen via grant number 11C5720N.
K.M. is supported by King's College London through a Postgraduate International Scholarship. M.S. is supported in part by the Science and Technology Facility Council (STFC), United Kingdom, under the research grant ST/P000258/1.

\bibliography{references}

\end{document}